\documentclass[%
 reprint,
 amsmath,amssymb,
 aps,
]{revtex4-1}

\usepackage{graphicx}
\usepackage{dcolumn}
\usepackage{bm}
\usepackage{xspace}
\usepackage{multirow}
\usepackage{booktabs}
\usepackage{xcolor}
\usepackage{xfrac}
\usepackage[caption=false]{subfig}
\graphicspath{{./plots/}}

\usepackage{hyperref}
\hypersetup{
    colorlinks=true,
    linkcolor=blue,
    filecolor=blue,      
    urlcolor=blue,
    pdftitle={Overleaf Example},
    pdfpagemode=FullScreen,
    }

\usepackage{units}
\usepackage{comment}

\def\sib{{\textsc{Sibyll}\,2.3c}\xspace}

\def\qgs{\textsc{QGSJet}-II.04\xspace}

\def\epos{\textsc{EPOS-LHC}\,\xspace}

\newcommand{\x}[1]{%
  {}$
  \kern-2\mathsurround 
  $
  \binoppenalty10000 \relpenalty10000 #1
  {}$
  \kern-2\mathsurround 
  $
}

\begin{document}


\title{$P_{\rm tail}^{\alpha}$: a high resolution gamma/hadron and composition discriminant variable for Water-Cherenkov Detector cosmic-ray observatories}

\author{Ruben Concei\c{c}\~{a}o}
\address{Laborat\'{o}rio de Instrumenta\c{c}\~{a}o e F\'{i}sica Experimental de Part\'{i}culas (LIP) - Lisbon, Av.\ Prof.\ Gama Pinto 2, 1649-003 Lisbon, Portugal}
\address{Instituto Superior T\'ecnico (IST), Universidade de Lisboa, Av.\ Rovisco Pais 1, 1049-001 Lisbon, Portugal}

\author{Pedro J. Costa}
\address{Laborat\'{o}rio de Instrumenta\c{c}\~{a}o e F\'{i}sica Experimental de Part\'{i}culas (LIP) - Lisbon, Av.\ Prof.\ Gama Pinto 2, 1649-003 Lisbon, Portugal}
\address{Instituto Superior T\'ecnico (IST), Universidade de Lisboa, Av.\ Rovisco Pais 1, 1049-001 Lisbon, Portugal}

\author{Lucio Gibilisco}
\email{gibilisc@lip.pt}
\address{Laborat\'{o}rio de Instrumenta\c{c}\~{a}o e F\'{i}sica Experimental de Part\'{i}culas (LIP) - Lisbon, Av.\ Prof.\ Gama Pinto 2, 1649-003 Lisbon, Portugal}
\address{Instituto Superior T\'ecnico (IST), Universidade de Lisboa, Av.\ Rovisco Pais 1, 1049-001 Lisbon, Portugal}

\author{M\'ario Pimenta}
\address{Laborat\'{o}rio de Instrumenta\c{c}\~{a}o e F\'{i}sica Experimental de Part\'{i}culas (LIP) - Lisbon, Av.\ Prof.\ Gama Pinto 2, 1649-003 Lisbon, Portugal}
\address{Instituto Superior T\'ecnico (IST), Universidade de Lisboa, Av.\ Rovisco Pais 1, 1049-001 Lisbon, Portugal}

\author{Bernardo Tom\'e}
\address{Laborat\'{o}rio de Instrumenta\c{c}\~{a}o e F\'{i}sica Experimental de Part\'{i}culas (LIP) - Lisbon, Av.\ Prof.\ Gama Pinto 2, 1649-003 Lisbon, Portugal}
\address{Instituto Superior T\'ecnico (IST), Universidade de Lisboa, Av.\ Rovisco Pais 1, 1049-001 Lisbon, Portugal}

\date{\today}

\begin{abstract}
The precise and efficient identification of the nature of the primary cosmic rays on an event-by-event basis stands as a fundamental aspiration for any cosmic ray observatory.
In particular, the detection and characterization of gamma ray events are challenged by their occurrence within an overwhelmingly greater flux of charged cosmic rays spanning several orders of magnitude. The intricacies of distinguishing between cosmic ray compositions and the inherent uncertainties associated with hadronic interactions present formidable challenges, which, if not properly addressed, can introduce significant sources of systematic errors.

This work introduces a novel composition discriminant variable, $P_{\rm tail}^{\alpha}$, which quantifies the number of  Water Cherenkov Detectors with a signal well above the mean signal observed in WCDs located at an equivalent distance from the shower core, in events with approximately the same energy at the ground. This new event variable is then shown to be, in the reconstructed energy range $10\,$TeV to $1.6\,$PeV, well correlated with the total number of muons that hit, in the same event, all the observatory stations located at a distance greater than $200\,{\rm m}$ from the shower core. The two variables should thus have similar efficiencies in the selection of high-purity gamma event samples and in the determination of the nature of charged cosmic ray events.

\end{abstract}

\pacs{Valid PACS appear here}
\maketitle


\section{Introduction}
\label{sec:intro}

The selection, with good efficiency and high purity, of highly energetic gamma rays or the determination of the nature of the charged cosmic rays is one of the major challenges for cosmic ray and gamma ray experiments. 

The direct detection of neutral and charged cosmic rays by high-altitude balloons or satellites is excluded at high energies (above tens of TeV for gamma rays and thousands of TeV for charged cosmic rays) due to the scarcity of such particles and the limited detection area of such detectors (typically a few ${\rm m^2}$) \cite{BLUMER2009293}. Thus, the only viable option is indirect detection, achieved by measuring the longitudinal development of the Extensive Air Shower (EAS) produced by the interaction of these particles in the Earth's atmosphere \cite{CTAConsortium:2013ofs,HESS:2015cyv}, or by studying the distribution of the EAS particles that reach the ground \cite{Auger,KASCADE:2003swk}.

Several different experimental methods and discriminant variables have been developed to select gamma-ray events from the huge hadronic background and to discriminate between showers that might have been produced by different atomic nuclei (typically from hydrogen to iron) \cite{Tian:2023lgr,ArteagaVelazquez:2023wzt}. No unique or perfect solution exists, although, above a few TeV, the direct measurement of the number of muons arriving at the ground is widely accepted as the best possible discriminator variable and has indeed allowed the detection of gamma rays with energies up to the PeV by the LHAASO collaboration~\cite{Peta_gamma,LHAASO_PeV}. More recently, a new gamma/hadron discriminating variable, $LCm$, based on the measurement of the azimuthal non-uniformity of the particle distributions at the ground in Water Cerenkov Detectors (WCD) arrays, was introduced \cite{LCm} and, through simulations, it has been claimed that it might reach equivalent background rejection factors of about $10^4$ at energies about $1\,$PeV~\cite{LCm_1PEV}. The latter quantity has, however, shown limited discrimination power for composition and hadronic interactions studies.

In this article, we introduce a novel variable denoted as $P_{\rm{tail}}^{\alpha}$ designed for WCD ground arrays. By focusing on events falling within a specific energy range at the ground, we construct distributions of signals across stations, categorized into discrete distance bins from the shower core. These distributions can be either derived from available data or generated through simulation when data is lacking. $P_{\text{tail}}^{\alpha}$ provides a quantitative measure on an event-by-event basis, indicating the number of stations exhibiting signals within the upper tail of these signal distributions.
The rationale behind this variable, inspired by a method developed by the IceTop/IceCube collaboration~\cite{icetop}, lies in the observation that in events with comparable reconstructed energy, the signal recorded by the WCD stations tends to be higher when struck by energetic sub-showers. These sub-showers, composed of muons and highly energetic electromagnetic particles, serve as a distinct signature of hadronically-induced showers \cite{HAWC:2022hny}.

The manuscript is organised as follows: in Section \ref{sec:simulation}, all the simulation sets are described; in Section \ref{sec:Ptail}, the new variable, $P_{\rm tail}^{\alpha}$, is introduced; in Section \ref{sec:PtailNmu}, the correlation of this new variable with the number of muons that hit the WCD stations in gamma, proton or iron events with reconstructed energy between $10\,$TeV and $1.6\,$PeV is analysed; in Section \ref{sec:discrimination}, the efficiency of this new variable to select high purity gamma event samples, as well as to determine the nature of charged cosmic rays events is reported; finally, in Section \ref{sec:conclusions}, the use of this new variable in the present and future large ground-array gamma-ray observatories is discussed.
 
\section{Simulation framework}
\label{sec:simulation}

CORSIKA (version 7.5600) \cite{CORSIKA} was used to simulate gamma-ray, proton-induced and iron-induced vertical ($\theta = 0^\circ$) showers assuming an observatory altitude of $5200\,$m a.s.l. The simulated shower energy ranged from $10\,$TeV up to $1.6\,$PeV, being generated with an $E^{-1}$ energy spectrum. In order to realistically replicate the $E^{-2}$ ($E^{-3}$) flux of gamma rays (charged cosmic rays), a further $E^{-1}$ ($E^{-2}$) weight on the simulated energy of the events has then been added in the analysis. FLUKA v~\cite{fluka,fluka2} and \qgs \cite{qgs} were used as hadronic interaction models for low- and high-energy interactions, respectively.

A ground detector array was emulated by a 2D-histogram, each cell representing a station with an area of $\approx 12\,{\rm m^2}$. The stations were arranged to cover a circular surface of $\sim 1\,{\rm km}$ radius with a uniform fill factor ($FF$) -- defined as the ratio between the instrumented area and the shower collection area. Different fill factors were obtained by masking the 2D-histogram with a regular pattern.

The signal in each station was estimated as the sum of the expected signals due to the particles hitting the station, using dedicated parameterizations as a function of the particle energy for protons, muons and electrons/gammas. These curves were obtained by injecting vertical particles sampled uniformly on top of a \emph{Mercedes} Cherenkov detector station~\cite{Mercedes}, a single-layer, small \footnote{$2\,{\rm m}$ radius, $1.7\,{\rm m}$ water height} WCD with $3$ PMTs arranged in a $120^{\circ}$ star configuration at its bottom.
This procedure was used to estimate the signal deposited by the Vertical Equivalent Muon (VEM) in the \emph{Mercedes} WCD as well. It was found that $1\,{\rm VEM}\simeq 244\,{\rm photoelectrons}$.

The parameterizations were built for the mean signal in the station and the signal distribution standard deviation. Through the use of these two numbers, it was possible to emulate the fluctuations in the signal response of the WCDs due to the stochastic processes of particle interactions and light collection. Additionally, for muons, the fluctuation in their tracklength due to geometry variations was included as well. This was achieved using the distribution of the muons taken from proton-induced shower simulations run over a Geant4 simulation, which provided the geometry of the WCD array and stations.

To mimic realistic experimental conditions, a basic energy reconstruction method was employed. Initially, a power law fit was applied to correlate the simulated energy $E_0$ with the total electromagnetic signal $S_{\rm em}$ measured by the array, starting from $40\,{\rm m}$ away from the shower core. Such calibration was used to reconstruct the primary energy.\\
The events where thus divided in bins of reconstructed energy ($E_{\rm rec}$) ranging from $10\,{\rm TeV}$ to $1.6\,{\rm PeV}$, each bin having a logarithmic width of $0.2$. This method allowed the comparison of showers with similar total signal at the ground, regardless of the primary particle.\\
For the generation of all figures in this article (with the exception of Figure \ref{fig:Nmures}), a bin of events with $E_{rec}$ ranging from $100\,{\rm TeV}$ to $\sim160\,{\rm TeV}$ has been chosen. This energy interval will be hereafter denoted as ``around $100\,{\rm TeV}$'' for the sake of brevity.

Additionally, the shower core reconstruction was simulated by introducing a Gaussian smearing of $5\,$m to the estimated shower core position. This approach is conservative within the energy range investigated in this study ($E_{0} \in [10, 1000]\,{\rm TeV}$) considering the studies reported in~\cite{SWGO,LHAASO,HOFMANN2020102479}. Additionally, further tests have been conducted with core reconstruction resolution values up to $20\,{\rm m}$, showing no degradation compared to the results presented here.

\section{The discriminant variable: $P_{\rm tail}^{\alpha}$}
\label{sec:Ptail}

The $P_{\rm tail}^{\alpha}$ variable is defined as:
\begin{equation}
P_{\rm tail}^{\alpha}=  \sum_{i}^{n} ({P_{{\rm tail}, i})}^{\alpha}  
\label{eq:Ptaila}
\end{equation}
Here, $P_{\text{tail},i}$ represents the probability that the signal observed in the $i^{th}$ station of the WCD cosmic ray observatory falls within the upper tail of the signal distribution. These observations occur in stations at a similar distance from the shower core in the shower's transverse plane and pertain to events with comparable energy at the ground. The variable $n$ indicates the count of active stations under consideration. The parameter $\alpha$ adjusts the significance of stations where $P_{\text{tail},i}$ approaches 1~\cite{P_alpha}. When $\alpha = 1$, $P_{\text{tail}}^{\alpha}$ equals the sum of probabilities across all individual stations.

To avoid the core region, where the signals, dominated by the electromagnetic component, are extremely high and even saturation on its measured values may occur, the stations located at distances to the shower core smaller than $200\,{\rm m}$ are discarded.

${P_{{\rm tail},i}}$ is  computed, in each event and for each station $i$ as:
\begin{equation}
{P_{{\rm tail},i}}=  C_{r_{i}}(S_{i}) 
\label{eq:Ptaili}
\end{equation}
where $S_{i}$ is the signal observed in the $i^{th}$ station of the event. The function $C_{r_{i}}$ represents the normalized cumulative distribution of signals detected within a circular ring situated in the shower's transverse plane, beginning at a distance $r_{i}$ from the shower axis and with a width of $10\,{\rm m}$. 

The cumulative distributions for each ring are constructed from a set of shower events with the same reconstructed energy.

\begin{figure}[!t]
\centering
\includegraphics[width=0.9\linewidth]{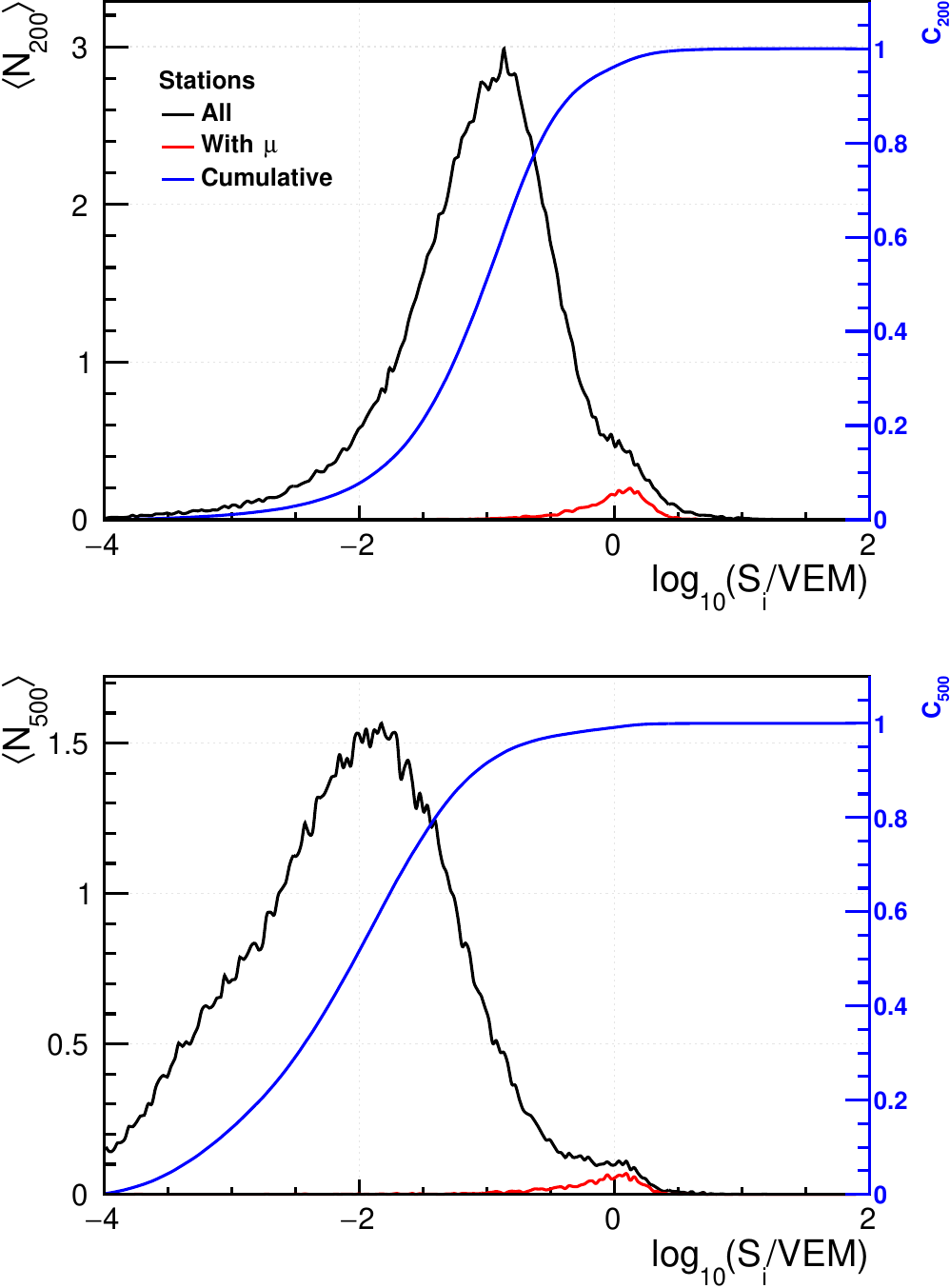}
\caption{\label{fig:cumulatives} Distributions of the total signal (black lines) in the stations in a $10\,{\rm m}$-wide ring at $200\,{\rm m}$ (top) and $500\,{\rm m}$ (bottom) from the shower core, with the respective cumulative distributions (blue lines), for $\mathcal{O}(10^3)$ proton showers with $E_{\rm rec}$ around $100\,{\rm TeV}$, measured in a ground array with fill factor $FF \simeq 12.5\%$. 
The red lines represent the distributions of the signal in the stations hit by muons.}
\end{figure}

As an example, the two distributions of the total signal and their cumulative distributions in the rings with a radius $r_{i}$ of $200\,{\rm m}$  and $500\,{\rm m}$, are shown in Figure \ref{fig:cumulatives}. 
These cumulative distributions are, as defined by Equation \ref{eq:Ptaili},  the functions ${P_{{\rm tail},i}}$ for the corresponding rings.
The signals of the stations hit by muons are also identified. These signals are, as expected, in the tail of the distributions. It is noteworthy that although these cumulative distributions were constructed using a specific high-energy hadronic interaction model, \qgs, a comparative assessment was undertaken using alternative models—namely, \epos and \sib. Remarkably, no significant disparities among the models were observed.

\begin{figure}[!t]
\centering
\includegraphics[width=0.9\linewidth]{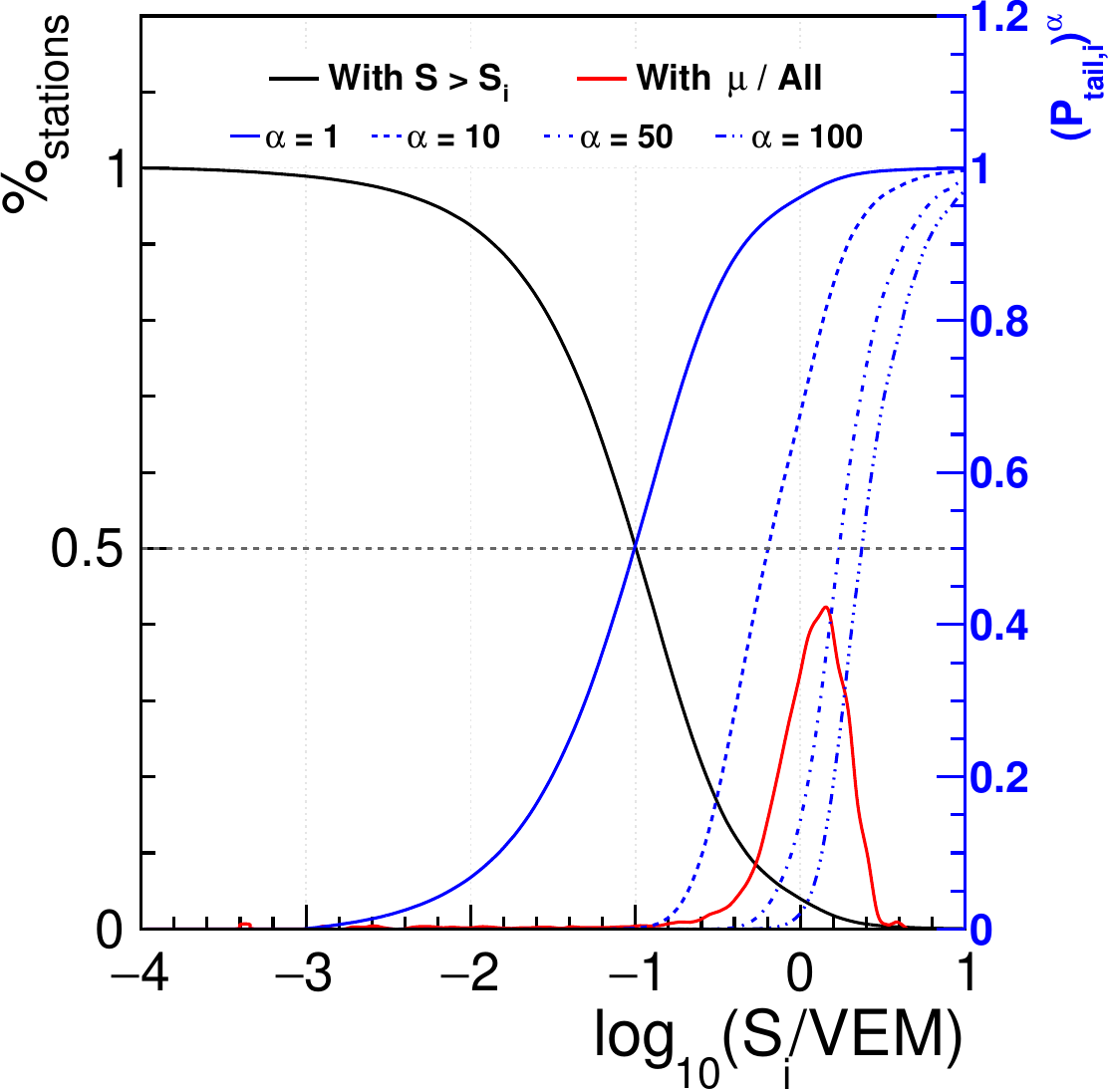}
\caption{\label{fig:Ptaili_alpha} 
Normalised number of stations of a ground array with a fill factor $FF \simeq 12.5\%$ situated in a $10\,{\rm m}$-wide ring at $300\,{\rm m}$  from the shower core, that have a signal higher than $S_{i}$, for $\mathcal{O}(10^3)$ proton showers with $E_{\rm rec}$ around $100\,{\rm TeV}$ (black curve).
The blue curves are the distributions of $({P_{{\rm tail},i})}^{\alpha}$ as a function of $S_{i}$ for $\alpha = 1, 10, 50, 100$ and the red line is the $\%$ of stations with a signal equal to $S_{i}$ that have been hit by muons.}
\end{figure}

For the sake of the readability of the plots in Figure \ref{fig:cumulatives} (and Figure \ref{fig:Ptaili_alpha} as well), the signal range shown extends down to $10^{-4}$ VEM, however, it should be noted that this choice does not aim at being a representation of a realistic signal threshold. In fact, we estimate that this threshold could be comfortably raised at least to $10^{-2}$ VEM without affecting the sensitivity of $P_{\rm tail}^{\alpha}$ to the high-signal tail. Any further discussion of the low-signal threshold is out of the scope of the present article and shall be assessed in a future study.

The effect of the $\alpha$ parameter is demonstrated in Figure~\ref{fig:Ptaili_alpha} for proton events with reconstructed energies around $100\,{\rm TeV}$ 
and considering the ring situated at $300\,{\rm m}$ from the shower core. The normalised number of stations that have a signal higher than $S_{i}$ are shown as a black line, while the functions $({P_{{\rm tail},i})}^{\alpha}$  as a function of the total signal, $S_{i}$, for $\alpha = 1, 10, 50, 100 $ are presented in blue. In red, the percentage of stations with a signal equal to $S_{i}$ that have been hit by muons is shown.
Note that, for $\alpha \sim  50 $ and $({P_{{\rm tail},i})}^{\alpha} = 0.5 $, half of the stations were hit at least by one muon.
Hereafter, for simplicity, the  $\alpha$ parameter is set to $50$.

To perform the analysis presented throughout the article, a fill factor $FF=12.5\%$ has been employed. A lower fill factor, $FF=5\%$, has been tested as well. Within the currently available statistics, no significant difference between the two configurations is found. Furthermore, the results shown in this article have been obtained using only vertical showers.

A study with inclined showers has been conducted, showing an overall increase in the $P_{\rm tail}^{\alpha}$ due to a greater absorption of the electromagnetic component of the shower, that leads to the presence of a higher ratio of stations with a high (muonic) signal and therefore a higher $P_{\rm tail,i}$. Through the tuning of the $\alpha$ parameter, the thorough exploration of which will be among the topics of future publications, a discrimination power equivalent to the one obtained with vertical showers can be achieved.

\section{Correlation of $P_{\rm tail}^{\alpha}$ with the total number of detected muons }
\label{sec:PtailNmu}

\begin{figure}[!t]
\centering
\includegraphics[width=0.95\linewidth]{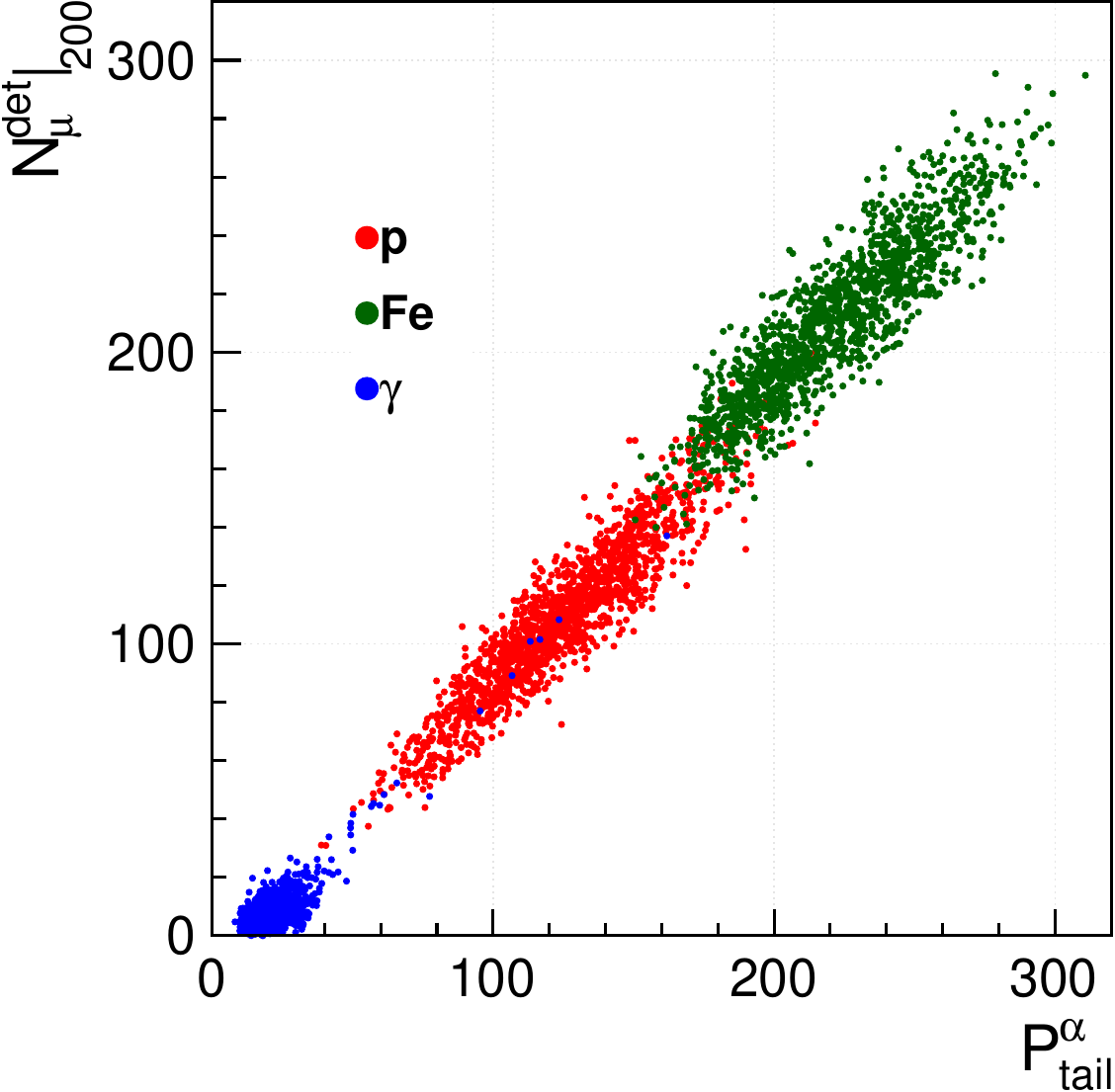}
\caption{\label{fig:nmuVptail} Correlation between the number of muons detected in an array with fill factor $FF \simeq 12.5\%$ at a distance from the shower core greater than $200\,{\rm m}$, $N_{\mu}^{\rm det}$, and $P_{\rm tail}^{\alpha}$ for gamma (blue), proton (red) and iron (green) events with reconstructed energy  around $100\,{\rm TeV}$. The energy was reconstructed under the assumption that each event was identified as a proton event.}
\end{figure}

The correlation of the new variable $P_{\rm tail}^{\alpha}$ with the total number of muons, $N_{\mu}^{\rm det}|_{200}$, that hit the WCD stations at a distance from the shower core greater than $200\,{\rm m}$ is shown in Figure \ref{fig:nmuVptail} using gamma, proton and iron samples with reconstructed energy around $100\,{\rm TeV}$ and an array with fill factor $FF \simeq 12.5\%$.

To a first order, the two variables exhibit an almost linear correlation, which allows the use of $P_{\rm tail}^{\alpha}$ to estimate, event by event, $N_{\mu}^{\rm det}|_{200}$, with high resolution and minimal bias.  
$N_{\mu}^{\rm det}|_{200}$ was estimated as:
\begin{equation}
 N_{\mu}^{\rm det \star}|_{200} = K_i  P_{\rm tail}^{\alpha},
\label{eq:Nmu_*}
\end{equation}
where $K_i$ are normalisation constants, each determined in the proton sample with the relevant energy range within the large spectrum of reconstructed energies (from ${10\,\rm TeV}$ to  $1.6\,{\rm PeV}$). The parameter $K_i$ varies between $0.85$ and $1.78$ across different energy ranges. These variations, for a fixed value of the $\alpha$ parameter across all energy ranges, can be attributed to differences in the size of the electromagnetic component in the stations and variations in the average number of muons in the stations hit by muons.

\begin{figure}[!t]
\centering
\includegraphics[width=0.95\linewidth]{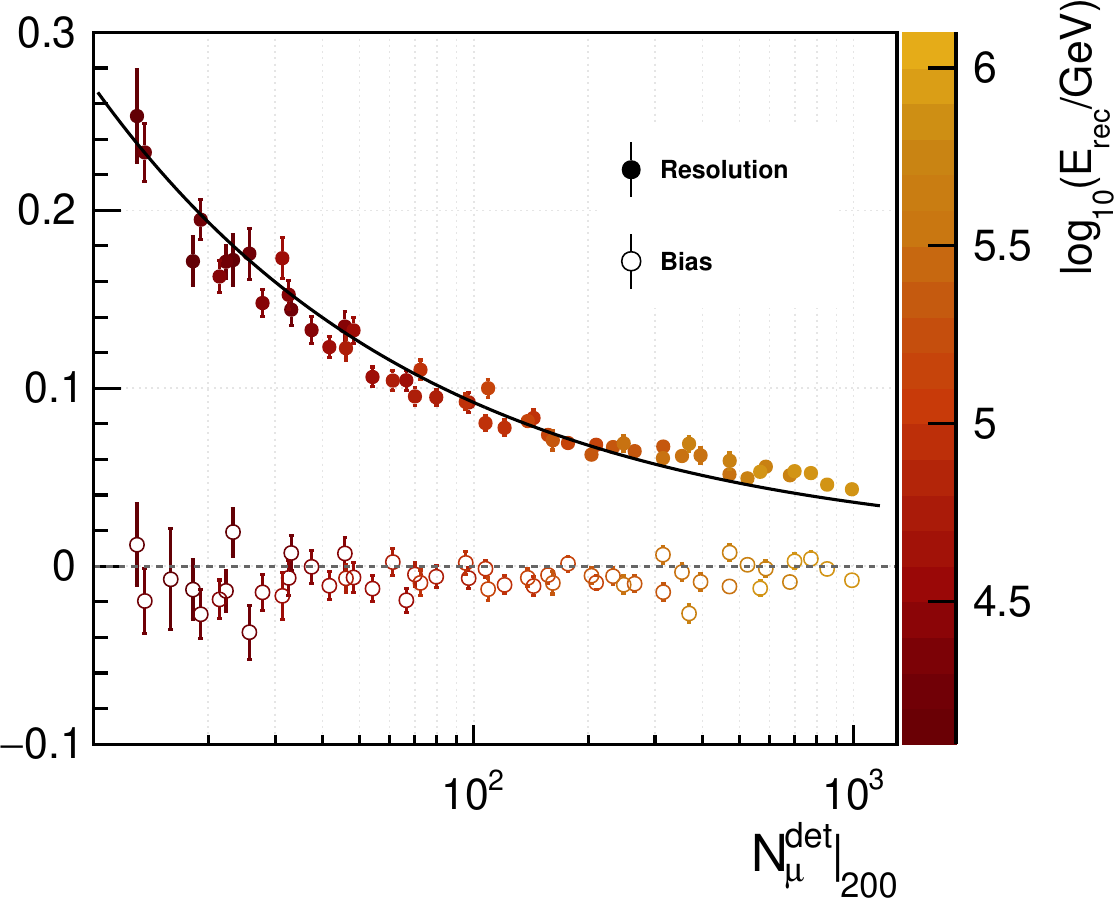}
\caption{\label{fig:Nmures} Resolution (full circles) and Bias  (open circles) of the estimation of the  number of muons that have hit the WCD stations at a distance from the shower core greater than $200\,{\rm m}$ from the measured value of $P_{\rm tail}^{\alpha}$ in an array with fill factor $FF \simeq 12.5\%$. 
See the color scale to identify the several $E_{\rm rec}$ samples used in the analysis.}
\end{figure}

The resolution on  $N_{\mu}^{\rm det \star}|_{200}$  was then assessed via the quantity: 
\begin{equation}
 \Delta N_{\mu}^{\rm det \star}|_{200} = \frac{ N_{\mu}^{\rm det}|_{200} -  N_{\mu}^{\rm det \star}|_{200}}{N_{\mu}^{\rm det }|_{200}}.
\label{eq:DNmu_*}
\end{equation}
Its absolute values and corresponding bias were computed from $10\,{\rm TeV}$ to $1.6\,{\rm PeV}$ and are summarised in Figure~\ref{fig:Nmures}, as a function of the total number of muons that hit the stations. 

The resolution was found to be essentially determined by the number of detected muons and is well described by the function (also represented in the same figure): 
\begin{equation}
\sigma_{\Delta N_{\mu}^{\rm det \star}|_{200}} = A + \frac{B}{\sqrt{N_{\mu}^{\rm det }|_{200}}  },
\label{eq:Par_DNmu_*}
\end{equation}
with $A \simeq 0.01 $ and $B \simeq 0.82$.

The bias was found to be below $ 2\%$.

To a second order, small systematic effects due to the differences in the shower development stage, namely on the maximum shower depth, $X_{\rm max}$, with different primary energy or primary nature, were found to contribute to the small observed bias. This effect is more evident when comparing proton and iron simulation sets with the same energy --  for instance, the red and green distributions of Figure~\ref{fig:nmuVptail}.
Nevertheless, the investigation into the influence of these factors and their potential mitigation strategies, as well as their utility in scrutinising various available hadronic interaction models within the same primary energy bin, lies beyond the scope of this current article and is planned to be explored in a forthcoming publication.


\section{$\gamma / {\rm h}$ and composition discrimination}
\label{sec:discrimination}

The good correlation between $P_{\rm tail}^{\alpha}$ and $N_{\mu}$ reported in the previous section indicates that the effectiveness of both variables in $\gamma / {\rm h}$ and composition discrimination should be similar. 
 
\begin{figure}[!t]
\centering
\includegraphics[width=0.95\linewidth]{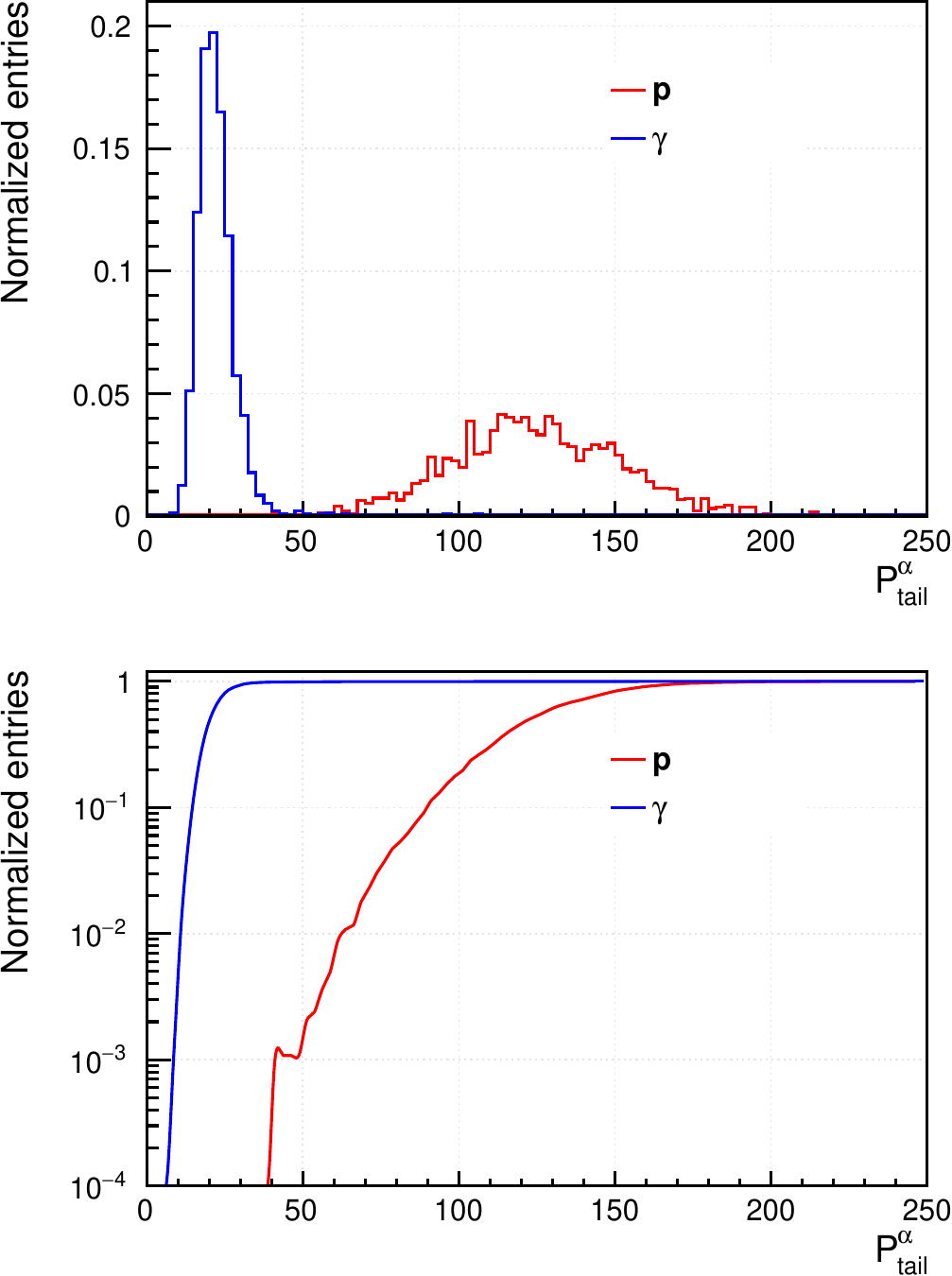}
\caption{\label{fig:g-p}
Distributions (top) of $P_{\rm tail}^{\alpha}$ for $\mathcal{O}(10^3)$ gamma events (blue line) with  energies  of around $100\,{\rm TeV}$ and proton events (red line) with similar energies at ground. On the bottom, the respective cumulative distributions are shown.}
\end{figure}

The $P_{\rm tail}^{\alpha}$ distributions, as well as their cumulative distributions, are shown in Figure \ref{fig:g-p} for gamma showers (blue lines) and proton showers (red lines) with reconstructed energy around $100\,{\rm TeV}$, considering an array with $FF \simeq 12.5\%$.
Within the statistics \footnote{$\mathcal{O}(10^3)$ events for proton showers}, proton rejection factors better than $10^{3}$ are achieved at a gamma efficiency close to $90\%$.
Such value can be compared to the hadron rejection factor achieved by LHAASO at $100\,{\rm TeV}$ through the ratio of the number of muons over the number of the electrons in the shower \cite{LHAASO:2024zug}, reportedly better than $1.5 \times 10^4$. Similarly, through the usage of the $PINCness$ parameter, the HAWC collaboration claims to reduce the fraction of gamma-ray showers mistakenly rejected to $\sim 4\%$ \cite{Abeysekara_2017}.

\begin{figure}[!t]
\centering
\includegraphics[width=0.95\linewidth]{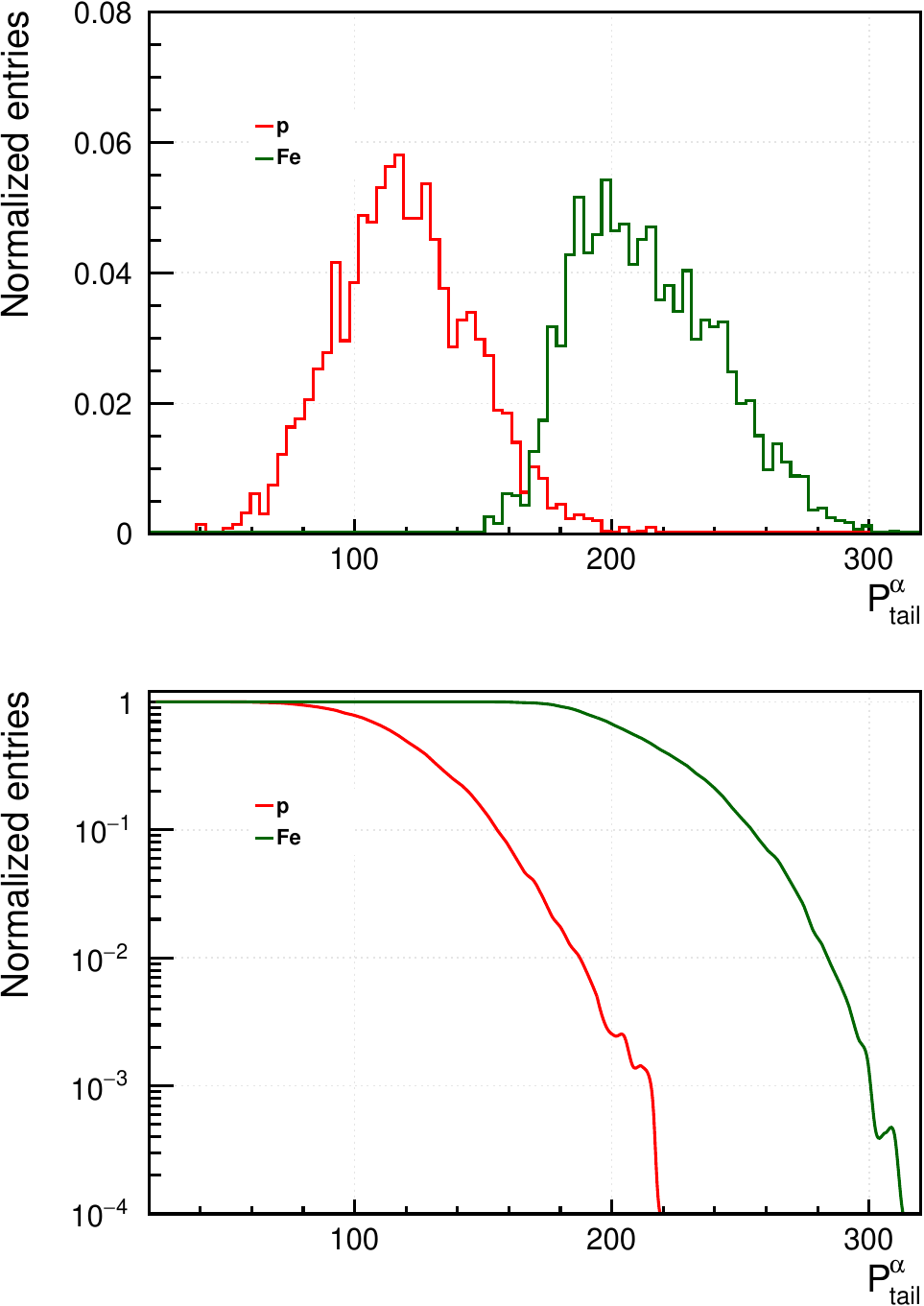}
\caption{\label{fig:p-Fe} 
Distributions (top) of $P_{\rm tail}^{\alpha}$ for $\mathcal{O}(10^3)$ proton events (red line) and iron events with reconstructed energy around $100\,{\rm TeV}$. On the bottom, the respective cumulative distributions are shown.}
\end{figure}

The $P_{\rm tail}^{\alpha}$ distributions as well as their cumulative distributions are shown in Figure~\ref{fig:p-Fe} for proton showers (red lines) with reconstructed energy around $100\,{\rm TeV}$ and iron showers (green lines) with similar reconstructed energy at the ground, considering an array with $FF \simeq 12.5\%$.

%
%

A good separation between the proton and iron distributions was observed and quantified using the selection efficiency of high-purity samples.
Fractions of protons as low as $\sim 1.26\times 10^{-2}$ are achieved at iron efficiencies close to $90\%$. Conversely, in the case of selecting high-purity proton samples, the fraction of iron obtained at proton efficiencies close to $90\%$ is $\sim 2.6 \times 10^{-3}$.


For reference, these discrimination, rejection and efficiency values may be compared with what would be achieved considering ideal muon detectors and using as a discriminant variable the number of muons, $N_{\mu}^{\rm det}$, that hit the detectors' surface area.

\begin{figure}[!t]
\centering
\includegraphics[width=0.95\linewidth]{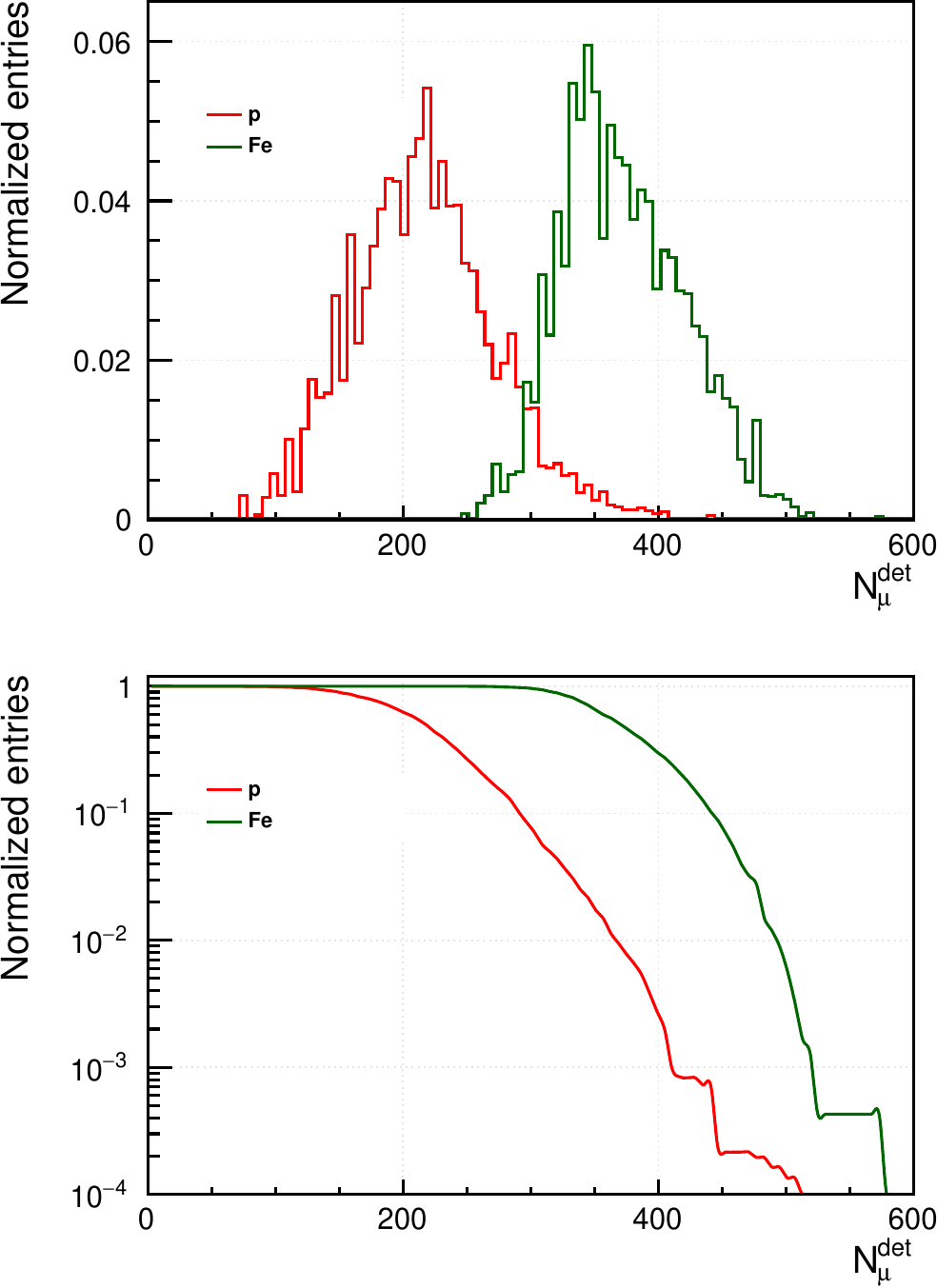}
\caption{\label{fig:p-Fe_Nmu} Distributions (top) of $N_{\mu}^{\rm det}$ for $\mathcal{O}(10^3)$ proton events (red line) and iron events (green line) with reconstructed energy of around $100\,{\rm TeV}$. On the bottom, the respective cumulative distributions are shown.}
\end{figure}

The $N_{\mu}^{\rm det}$ distributions, as well as their cumulative distributions, are shown in Figure \ref{fig:p-Fe_Nmu} for proton- (red lines) and iron-induced showers (green lines) with reconstructed energy around $100\,{\rm TeV}$, considering an array with $FF \simeq 12.5\%$.

Once again, we observe a notable distinction between the proton and iron distributions. Specifically, we obtain proton residual rates of approximately $5.0\times 10^{-2}$ at $90\%$. Vice versa, an iron residual rate close to $2.22\times 10^{-2}$ is achieved at proton efficiencies close to $90\%$.


This slightly worse result of $N_{\mu}^{\rm det}$ with respect to $P_{\rm tail}^{\alpha}$ can be understood noting that the latter is sensitive to the contributions of all the high energy secondary particles (muons, photons or electrons) coming from the interaction or the decay of hadronic particles produced in the event shower development, and not only to the muon component.

\section{Discussion and Conclusions}
\label{sec:conclusions}

The hadronic component of the EAS initiated by high-energy gammas, protons, or other nuclei is the main driver of the differences observed in their development. Namely, the decay of highly energetic $\pi$ and $k$-mesons produce muons that subsenquently arrive at the ground, but also, in the case of high-altitude observatories, high energy electromagnetic sub-clusters. 

The direct measurement of the number of muons arriving at the ground is, thus, an excellent gamma/hadron and composition discriminant variable. However, such measurements for EAS that are not highly inclined ($\theta < 60^{\circ}$) imply the shielding of the detectors from the huge EAS  electromagnetic component, usually implemented through the use of underground muon arrays. Such endeavor frequently incurs significant expenses and often exceeds the financial resources.

In this article, a new discriminant variable, $P_{\rm tail}^{\alpha}$, designed for WCD observatories, is introduced and discussed. This variable is easily built from the total signal measured in the array detectors. It is highly correlated with the total number of muons that would hit muon detectors with the same surface area and the contribution of the highly energetic electromagnetic sub-clusters. In this way, not surprisingly, the level of $\gamma / {\rm h}$  and composition discrimination of both variables was found to be similar, with the new variable being slightly better based on the selection efficiency of high-purity proton or iron samples. Furthermore, the resolution on the reconstruction, event by event, of the number of muons obtained from  $P_{\rm tail}^{\alpha}$ is just a function of the same number of muons and is about $10\%$ for $100$ predicted muons.

A percentage of approximately $10\%$ is less than half of the sigma of the distributions of the number of muons arriving at the detectors in an observatory with a fill factor of $12.5 \%$ for showers induced by $100\,{\rm TeV}$ protons. This is a comfortable operational region.

On the other hand, the number of muons is roughly directly proportional to the shower energy $E_0$ and the array fill factor. So, as a rule of thumb, resolutions of about $10\%$ on the number of muons hitting the detector are expected whenever $FF \times E_0 \sim 10$, which means that, in order to work at energies of about $1\,{\rm PeV}$ ($10\,{\rm TeV}$), fill factors of a few percent ($\sim 100\%$) are needed.

One should also emphasize that this new variable can be built from distributions directly measured in real data. In this case, $\alpha$ can be defined using the bump generated by stations with muons, as seen in Figure~\ref{fig:cumulatives}. Hence, its use as a discriminant variable for the different types of primaries is essentially independent of the choice of a given hadronic interaction model in simulations, which is usually one of the main sources of systematic errors.

The incorporation of this novel variable in conjunction with the established gamma/hadron discrimination variable $LCm$~\cite{LCm} holds significant promise. Its use and potential adaptation -- such as fine-tuning of the $\alpha$ parameter and the exclusion zone radius near the core, as discussed in Section~\ref{sec:Ptail} -- are subjects of active investigation for both present and future array observatories, including projects like SWGO~\cite{SWGO}. Furthermore, its applicability at higher energy regimes, extending up to the Auger Infill energy range (approximately $10^{17}\,{\rm eV}$), is currently being explored and will serve as the focal point of forthcoming publications.

\section*{Acknowledgments}
We would like to thank Ulisses Barres de Almeida, Antonio Bueno, Alessandro De Angelis, Jakub V\'icha and Alan Watson for carefully reading the manuscript.
This work has been financed by national funds through FCT - Fundação para a Ciência e a Tecnologia, I.P., under project PTDC/FIS-PAR/4300/2020.
L.~G. is grateful for the financial support by FCT under PRT/BD/154192/2022.
P.~C. is grateful for the financial support by FCT under UI/BD/153576/2022.

\bibliography{references}

\end{document}